\documentclass[aps,prl,preprint,floatfix,superscriptaddress,showpacs]{revtex4}

\usepackage{graphicx}
\usepackage{dcolumn}
\usepackage{ulem}
\usepackage{soul}

\begin{document}

\newcommand{\el}{\mbox{${\rm e^{-}}$ }}
\newcommand{\ps}{\mbox{${\rm e^{+}}$ }}

\title{The cosmic-ray positron energy spectrum measured by PAMELA}

\author{O. Adriani}
\affiliation{University of Florence, Department of Physics,  
 I-50019 Sesto Fiorentino, Florence, Italy}
\affiliation{INFN, Sezione di Florence,  
 I-50019 Sesto Fiorentino, Florence, Italy}
\author{G. C. Barbarino}
\affiliation{University of Naples ``Federico II'', Department of
Physics, I-80126 Naples, Italy}
\affiliation{INFN, Sezione di Naples,  I-80126 Naples, Italy}
\author{G. A. Bazilevskaya}
\affiliation{Lebedev Physical Institute, RU-119991
Moscow, Russia}
\author{R. Bellotti}
\affiliation{University of Bari, Department of Physics, I-70126 Bari, Italy}
\affiliation{INFN, Sezione di Bari, I-70126 Bari, Italy}
\author{A. Bianco}
\affiliation{INFN, Sezione di Trieste, I-34149
Trieste, Italy}
\affiliation{University of Trieste, Department of Physics, 
 I-34147 Trieste, Italy}
\author{M. Boezio}
\affiliation{INFN, Sezione di Trieste, I-34149
Trieste, Italy}
\author{E. A. Bogomolov}
\affiliation{Ioffe Physical Technical Institute,  RU-194021 St. 
Petersburg, Russia}
\author{M. Bongi}
\affiliation{University of Florence, Department of Physics,  
 I-50019 Sesto Fiorentino, Florence, Italy}
\affiliation{INFN, Sezione di Florence,  
 I-50019 Sesto Fiorentino, Florence, Italy}
\author{V. Bonvicini}
\affiliation{INFN, Sezione di Trieste,  I-34149
Trieste, Italy}
\author{S. Bottai}
\affiliation{INFN, Sezione di Florence,  
 I-50019 Sesto Fiorentino, Florence, Italy}
\author{A. Bruno}
\affiliation{University of Bari, Department of Physics, I-70126 Bari,
Italy} 
\affiliation{INFN, Sezione di Bari, I-70126 Bari, Italy}
\author{F. Cafagna}
\affiliation{INFN, Sezione di Bari, I-70126 Bari, Italy}
\author{D. Campana}
\affiliation{INFN, Sezione di Naples,  I-80126 Naples, Italy}
\author{R. Carbone}
\affiliation{INFN, Sezione di Trieste,  I-34149
Trieste, Italy}
\author{P. Carlson}
\affiliation{KTH Royal Institute of Technology, Department of Physics, and the Oskar Klein Centre for
Cosmoparticle Physics, AlbaNova University Centre, SE-10691 Stockholm,
Sweden}
\author{M. Casolino}
\affiliation{INFN, Sezione di Rome ``Tor Vergata'', I-00133 Rome, Italy}
\author{G. Castellini}
\affiliation{ IFAC,  I-50019 Sesto Fiorentino,
Florence, Italy}
\author{C. De Donato}
\affiliation{INFN, Sezione di Rome ``Tor Vergata'', I-00133 Rome, Italy}
\author{C. De Santis}
\affiliation{INFN, Sezione di Rome ``Tor Vergata'', I-00133 Rome, Italy}
\affiliation{University of Rome ``Tor Vergata'', Department of
Physics,  I-00133 Rome, Italy}
\author{N. De Simone}
\affiliation{INFN, Sezione di Rome ``Tor Vergata'', I-00133 Rome, Italy}
\author{V. Di Felice}
\affiliation{INFN, Sezione di Rome ``Tor Vergata'', I-00133 Rome, Italy}
\author{V. Formato}
\affiliation{INFN, Sezione di Trieste, I-34149
Trieste, Italy}
\affiliation{University of Trieste, Department of Physics, 
I-34147 Trieste, Italy}
\author{A. M. Galper}
\affiliation{National Research Nuclear University MEPhI,  RU-11540
Moscow, Russia}  
\author{A. V. Karelin}
\affiliation{National Research Nuclear University MEPhI,  RU-115409
Moscow, Russia}
\author{S. V. Koldashov}
\affiliation{National Research Nuclear University MEPhI,  RU-115409
Moscow, Russia}  
\author{S. A. Koldobskiy}
\affiliation{National Research Nuclear University MEPhI,  RU-115409
Moscow, Russia}  
\author{S. Y. Krutkov}
\affiliation{Ioffe Physical Technical Institute,  RU-194021 St. 
Petersburg, Russia}
\author{A. N. Kvashnin}
\affiliation{Lebedev Physical Institute, RU-119991
Moscow, Russia}
\author{A. Leonov}
\affiliation{National Research Nuclear University MEPhI,  RU-115409
Moscow, Russia}  
\author{V. Malakhov}
\affiliation{National Research Nuclear University MEPhI,  RU-115409
Moscow, Russia}  
\author{L. Marcelli}
\affiliation{University of Rome ``Tor Vergata'', Department of
Physics,  I-00133 Rome, Italy}
\author{M. Martucci}
\affiliation{University of Rome ``Tor Vergata'', Department of
Physics,  I-00133 Rome, Italy} 
\affiliation{INFN, Laboratori Nazionali di Frascati, Via Enrico Fermi 40,
I-00044 Frascati, Italy}
\author{A. G. Mayorov}
\affiliation{National Research Nuclear University MEPhI,  RU-115409
Moscow, Russia}
\author{W. Menn}
\affiliation{Universit\"{a}t Siegen, Department of Physics,
D-57068 Siegen, Germany}
\author{M. Merg\'{e}}
\affiliation{INFN, Sezione di Rome ``Tor Vergata'', I-00133 Rome, Italy}
\affiliation{University of Rome ``Tor Vergata'', Department of
Physics,  I-00133 Rome, Italy} 
\author{V. V. Mikhailov}
\affiliation{National Research Nuclear University MEPhI,  RU-115409
Moscow, Russia}  
\author{E. Mocchiutti}
\affiliation{INFN, Sezione di Trieste,  I-34149
Trieste, Italy}
\author{A. Monaco}
\affiliation{University of Bari, Department of Physics, I-70126 Bari, Italy}
\affiliation{INFN, Sezione di Bari, I-70126 Bari, Italy}
\author{N. Mori}
\affiliation{INFN, Sezione di Florence,  
 I-50019 Sesto Fiorentino, Florence, Italy}
\author{R. Munini}
\affiliation{INFN, Sezione di Trieste, I-34149
Trieste, Italy}
\affiliation{University of Trieste, Department of Physics, 
I-34147 Trieste, Italy}
\author{G. Osteria}
\affiliation{INFN, Sezione di Naples,  I-80126 Naples, Italy}
\author{F. Palma}
\affiliation{INFN, Sezione di Rome ``Tor Vergata'', I-00133 Rome, Italy}
\affiliation{University of Rome ``Tor Vergata'', Department of
Physics,  I-00133 Rome, Italy} 
\author{P. Papini}
\affiliation{INFN, Sezione di Florence,  
 I-50019 Sesto Fiorentino, Florence, Italy}
\author{M. Pearce}
\affiliation{KTH Royal Institute of Technology, Department of Physics, and the Oskar Klein Centre for
Cosmoparticle Physics, AlbaNova University Centre, SE-10691 Stockholm,
Sweden}
\author{P. Picozza}
\affiliation{INFN, Sezione di Rome ``Tor Vergata'', I-00133 Rome, Italy}
\affiliation{University of Rome ``Tor Vergata'', Department of
Physics,  I-00133 Rome, Italy} 
\author{C. Pizzolotto}
\altaffiliation[Previously at ]{INFN, Sezione di Trieste, I-34149
Trieste, Italy}
\affiliation{INFN, Sezione di Perugia, I-06123 Perugia, Italy}
\affiliation{Agenzia Spaziale Italiana (ASI) Science Data Center, I-00044 
Frascati, Italy}
\author{M. Ricci}
\affiliation{INFN, Laboratori Nazionali di Frascati, Via Enrico Fermi 40,
I-00044 Frascati, Italy}
\author{S. B. Ricciarini}
\affiliation{INFN, Sezione di Florence, 
 I-50019 Sesto Fiorentino, Florence, Italy}
\author{L. Rossetto}
\affiliation{KTH Royal Institute of Technology, Department of Physics, and the Oskar Klein Centre for
Cosmoparticle Physics, AlbaNova University Centre, SE-10691 Stockholm,
Sweden}
\author{R. Sarkar}
\affiliation{INFN, Sezione di Trieste, I-34149
Trieste, Italy}
\author{V. Scotti}
\affiliation{University of Naples ``Federico II'', Department of
Physics, I-80126 Naples, Italy}
\affiliation{INFN, Sezione di Naples,  I-80126 Naples, Italy}
\author{M. Simon}
\affiliation{Universit\"{a}t Siegen, Department of Physics,
D-57068 Siegen, Germany}
\author{R. Sparvoli}
\affiliation{INFN, Sezione di Rome ``Tor Vergata'', I-00133 Rome, Italy}
\affiliation{University of Rome ``Tor Vergata'', Department of
Physics,  I-00133 Rome, Italy} 
\author{P. Spillantini}
\affiliation{University of Florence, Department of Physics,  
 I-50019 Sesto Fiorentino, Florence, Italy}
\affiliation{INFN, Sezione di Florence,  
 I-50019 Sesto Fiorentino, Florence, Italy}
\author{S. J. Stochaj}
\affiliation{INFN, Sezione di Trieste,  I-34149
Trieste, Italy}
\author{J. C. Stockton}
\affiliation{INFN, Sezione di Trieste,  I-34149
Trieste, Italy}
\author{Y. I. Stozhkov}
\affiliation{Lebedev Physical Institute, RU-119991
Moscow, Russia}
\author{A. Vacchi}
\affiliation{INFN, Sezione di Trieste,  I-34149
Trieste, Italy}
\author{E. Vannuccini}
\affiliation{INFN, Sezione di Florence, 
 I-50019 Sesto Fiorentino, Florence, Italy}
\author{G. I. Vasilyev}
\affiliation{Ioffe Physical Technical Institute, RU-194021 St. 
Petersburg, Russia}
\author{S. A. Voronov}
\affiliation{National Research Nuclear University MEPhI,  RU-115409
Moscow, Russia}
\author{Y. T. Yurkin}
\affiliation{National Research Nuclear University MEPhI,  RU-115409
Moscow, Russia}  
\author{G. Zampa}
\affiliation{INFN, Sezione di Trieste,  I-34149
Trieste, Italy}
\author{N. Zampa}
\affiliation{INFN, Sezione di Trieste,  I-34149
Trieste, Italy}
\author{V. G. Zverev}
\affiliation{National Research Nuclear University MEPhI,  RU-115409
Moscow, Russia}  

\date{\today}
\begin{abstract}
Precision measurements of the positron component in the cosmic
radiation 
provide important information about the propagation of
cosmic rays and the nature of particle sources in
our Galaxy. 
The satellite-borne experiment PAMELA has been used to make a new
measurement of the cosmic-ray positron flux and fraction
that extends previously published measurements 
up to 300 GeV in kinetic energy. 
The combined
measurements of the cosmic-ray positron energy
spectrum and fraction provide a unique tool to constrain
interpretation models. 
During the recent solar minimum activity
period from July 2006 to December 2009
approximately 24500 positrons were observed. 
The results cannot be easily reconciled 
with purely secondary production and additional sources of
either astrophysical or exotic origin may be required.

\end{abstract}


\maketitle


Cosmic-ray positrons were first observed during pioneering experiments
in the sixties~\cite{des64} 
using balloon-borne magnetic spectrometers.
Positrons 
are a natural component of the cosmic radiation,  
produced in the interaction between  
cosmic rays and the interstellar matter. 
Since the first calculations of secondary positron
fluxes 
(e.g.,~\cite{pro82}) positrons 
have been shown to be 
extremely interesting for understanding 
the propagation mechanisms of cosmic rays. 
Furthermore, novel sources of primary cosmic-ray positrons of
either astrophysical or exotic origin can also be probed. 

Since July 2006, PAMELA (a Payload for
Antimatter Matter Exploration and Light-nuclei Astrophysics) has been 
measuring the antiparticle component of the cosmic radiation. 
A previous PAMELA measurement of the
positron fraction, the ratio of positron and electron fluxes: 
\mbox{$\phi$(e$^+$) / ($\phi$(e$^+$) + $\phi$(e$^-$))}, 
between 1.5 and 100 GeV~\cite{adr09b,adr10a} showed 
the first clear deviation
from secondary production models. 
Very recently, this was confirmed by the AMS collaboration that
presented results on the high-energy positron fraction~\cite{agu13} 
in excellent agreement with the PAMELA data.
A subsequent PAMELA measurement of
the cosmic-ray antiproton energy spectrum~\cite{adr10b} was found
to be consistent with 
expectations from secondary production calculations.  
In order to explain these results both 
dark matter and
astrophysical objects
(e.g., pulsars) 
have been proposed as positron sources
(e.g.,~\cite{boe09}). 
More than 20 years ago a positron excess at high
energy was  
postulated 
for the annihilation of dark matter particles in the
galactic halo (e.g.,~\cite{tyl89,kam91}). 
While extremely intriguing, such an explanation is challenged by 
the asymmetry between the leptonic (positrons)
and hadronic (antiprotons) PAMELA data.
A very high mass neutralino (e.g.,~\cite{cir08}) is required if this is the 
dominant dark matter species. The allowed supersymmetric parameter space 
does not favour this scenario. A dark matter contribution may
require pure leptonic annihilation 
channels (e.g.,~\cite{cir08}) or the introduction of a new dark sector
of forces (e.g.,~\cite{cho08}).
A contribution from 
pulsars would naturally increase the
positron and electron abundances (e.g.,~\cite{ato95}) 
without affecting the antiproton
component. Other astrophysical
models \cite{bla09b,fuj09,ahl09} have been proposed to explain the
PAMELA 
positron results with 
an, as yet unobserved, increase  
in the antiproton and secondary nuclei abundances predicted at high
energies ($\geq$100 GeV/n).
A detailed measurement of the positron energy
spectrum 
complements information from the positron fraction 
and provides stronger constraints
on theoretical models than possible from the positron fraction alone.

The PAMELA experiment \cite{pic07,boe09} comprises 
(from top to bottom): a time of flight system (TOF),
a magnetic spectrometer with silicon tracker planes,
an anticoincidence system (AC),
an electromagnetic imaging calorimeter,
a shower tail catcher scintillator and 
a neutron detector. These components
are housed inside a
pressurized container attached
to the Russian Resurs-DK1 satellite, which was launched on
June 15$^{{\rm th}}$ 2006. The orbital altitude varied between
350~km and 600~km at an inclination of 70$^\circ$.
Data presented here were acquired over the recent solar minimum activity 
period from July 2006 to December 2009
(1229 days), 
corresponding to more than $2 \times 10^9$ triggers.

Downward-going particles were selected using the ToF information. The 
time-of-flight resolution of 300~ps ensured that no contamination from
albedo particles remained in the  
selected sample. 
The ionization losses in the ToF scintillators and in the
silicon tracker layers 
were used to  
select minimum ionizing singly charged particles. 
Furthermore, multiply charged tracks were  
rejected by requiring no spurious signals in the ToF and AC
scintillators above the tracking system.  

Positrons were identified by combining information from
the different detector components. The misidentifications
of electrons and, in particular, protons are the largest sources of
background when estimating the positron signal. 
This can occur if the sign-of-charge is incorrectly assigned
(``spillover'' events) from  
the spectrometer data, or if 
electron- and proton-like interaction
patterns are confused in the calorimeter data.  
Using strict
selection criteria on the quality of the fitted tracks, the 
electron spillover background was estimated from flight 
data and simulation to be negligible below 
approximately 300 GeV. 
The proton background is much larger than the positron signal. The
proton-to-positron flux ratio increases from approximately 200 at 1
GV/c to approximately 2000 at 
100 GV/c. Robust positron identification is therefore required.
Positron identification is based on a combination of the calorimeter
shower topology and rigidity information from the tracking
system. In our first publications of the positron
component~\cite{adr09b,adr10a} a classical analysis was employed
applying strict criteria to this information. In this work the 
information was processed using a 
multivariate approach providing a significant increase in the positron
selection efficiency and a cross-check of the previously published
data. Specifically, the ``multilayer perceptron''
(MLP) network \cite{rum86} type of   
artificial neural network
\cite{boc04} implemented in the TMVA \cite{tmva} tool kit was used.
A set of 24 classification variables was chosen in order to fully 
represent the topology of the shower inside the calorimeter. The
analysis was performed in intervals of rigidity using events generated
by a Monte Carlo simulation of the PAMELA apparatus based on the 
GEANT4 code~\cite{ago03}. 
The samples of simulated
electrons and protons were divided into two parts. The first part was
used for 
the training of the multivariate algorithms while the second part was
used to test the
classifiers. Once the training and the testing
had been performed, it was possible to classify the negatively-charged
(mostly electrons) and
positively-charged (protons + positrons) particles selected from the
flight data.

\begin{figure}[h]
\begin{center}
\includegraphics[width=\columnwidth,height=0.82\textheight]{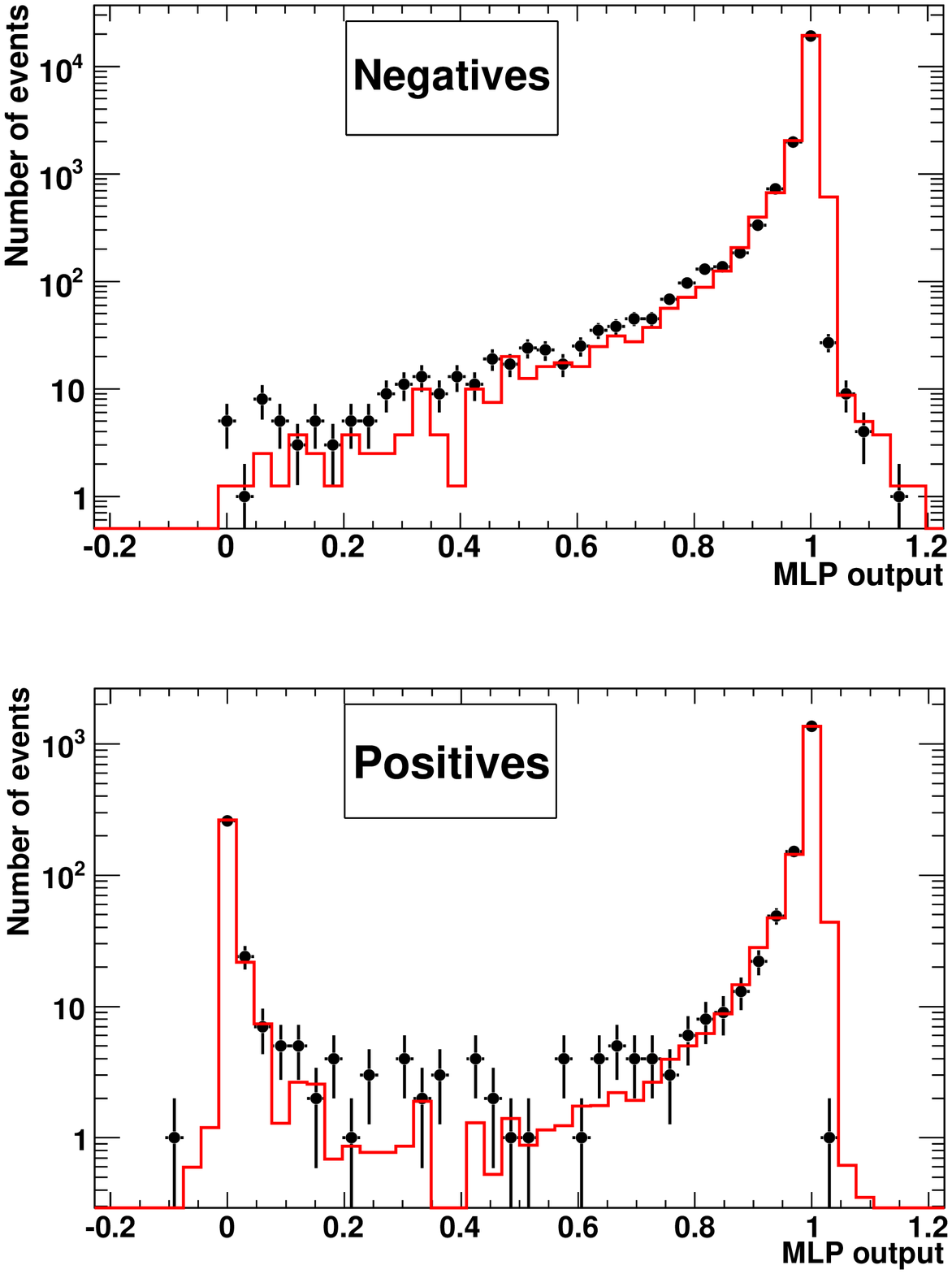}\\
\caption{The neural network output distributions for 
negatively-charged particles
(top panel) and 
positively-charged particles (bottom panel) for 
the rigidity bin 2.1-2.4 
GV/c. Top panel: distribution of simulated
electrons (solid red line) superimposed on the 
distribution of negatively-charged particles. Bottom panel: 
distribution for
positively-charged events (protons and positrons) fitted by 
the sum 
of the simulated electrons and protons distributions (solid red
line).}\label{fig1}
\end{center}
\end{figure}
Up to 20 GV/c, the number of electrons for each rigidity interval was
obtained by 
fitting the output distribution for negatively-charged particles 
with the 
distribution: $p\mathcal{F}_e$, where $\mathcal{F}_e$ is the output
distribution of simulated electrons and $p$ is a normalization
parameter resulting from the fit. Similarly, the
number of positrons was estimated from the fit of the output
distributions 
for positively charged particles using  
a mixture distribution of the neural network output for simulated
electrons and protons. Figure~\ref{fig1} shows the
neural output distributions  
for negatively-charged (upper panel)
and for positively-charged particles (lower
panel) for the rigidity interval 2.1-2.4 GV/c. 
A simple and high efficient event pre-selection 
was applied in order to have a reliable track reconstruction for 
energy and charge sign determination and to reduce the
proton contamination by about a factor 100 requiring a shower in the
calorimeter.  
In the bottom panel, events in
the left peak of the distribution correspond to the proton
residual contamination, while 
the peak to the right consists mostly of positron events.
Excellent agreement is found between the distribution for negative
events and the simulated electron distribution (solid red line in the
upper panel) and between the distribution for 
positive events
(protons and positrons) and the sum of the simulated electron and
proton distributions (solid red line in the lower panel). 
These results were cross-checked using another multivariate
classifier, ``random forest'' \cite{bre01}, 
a simulation with
enhanced $\pi^0$ production~\cite{ros12} and a classical
analysis~\cite{bia12} selecting positrons above 28~GeV. Each approach
produced consistent results. 
At rigidities higher than 20 GV/c, positrons (and electrons)
were selected by  
applying additional conditions based on calorimeter information 
to events yielding output 
values greater than 0.6 in Figure~\ref{fig1}.
A bootstrap procedure~\cite{efr93} was used to estimate 
the number of positrons and the positron fraction. For more details
on the positron analysis see~\cite{adr13b}.

The positron energy spectrum was derived by correcting the number of
positrons for 
selection efficiencies, live time and geometrical factor.
Efficiencies were estimated directly from flight data with the
exception of the track selection efficiency that was obtained from
Monte Carlo simulation (e.g., see~\cite{adr13a}). The live time 
was provided by 
an on-board clock that timed the periods during which the apparatus
was waiting for a trigger. The geometrical factor was estimated with
simulation to be
constant at 19.9 cm$^{2}$sr in the energy range of interest.
Positron energy spectra were obtained for different intervals of 
vertical geomagnetic cutoff,
estimated in
the St\"{o}rmer approximation~\cite{sto31}
using the satellite orbital
information. 
The 
energy spectra were unfolded using a Bayesian unfolding
procedure~\cite{dag95}. 
Spectra were combined accounting for
the proper live times and using only the fluxes at energies
that exceeded 1.3 times the maximum vertical geomagnetic cutoff at each
cutoff interval. 
The total systematic uncertainty on the flux was obtained by 
summing in quadrature the various systematic 
errors due to acceptance, efficiency
estimation and spectrum unfolding. A systematic
uncertainty on the overall flux estimation was derived by comparing the
electron energy spectra obtained using 
the calorimeter and the tracking information. The two sets of fluxes
differed by about 5\% at 2 GeV linearly increasing to 17\% at
100 GeV. Thus, the total systematic uncertainty on the flux was found to
vary from $\simeq 6\%$ at 2 GV to $\simeq 20\%$ above 100
GV.

\begin{figure}[ht]
\begin{center}
\includegraphics[width=\columnwidth]{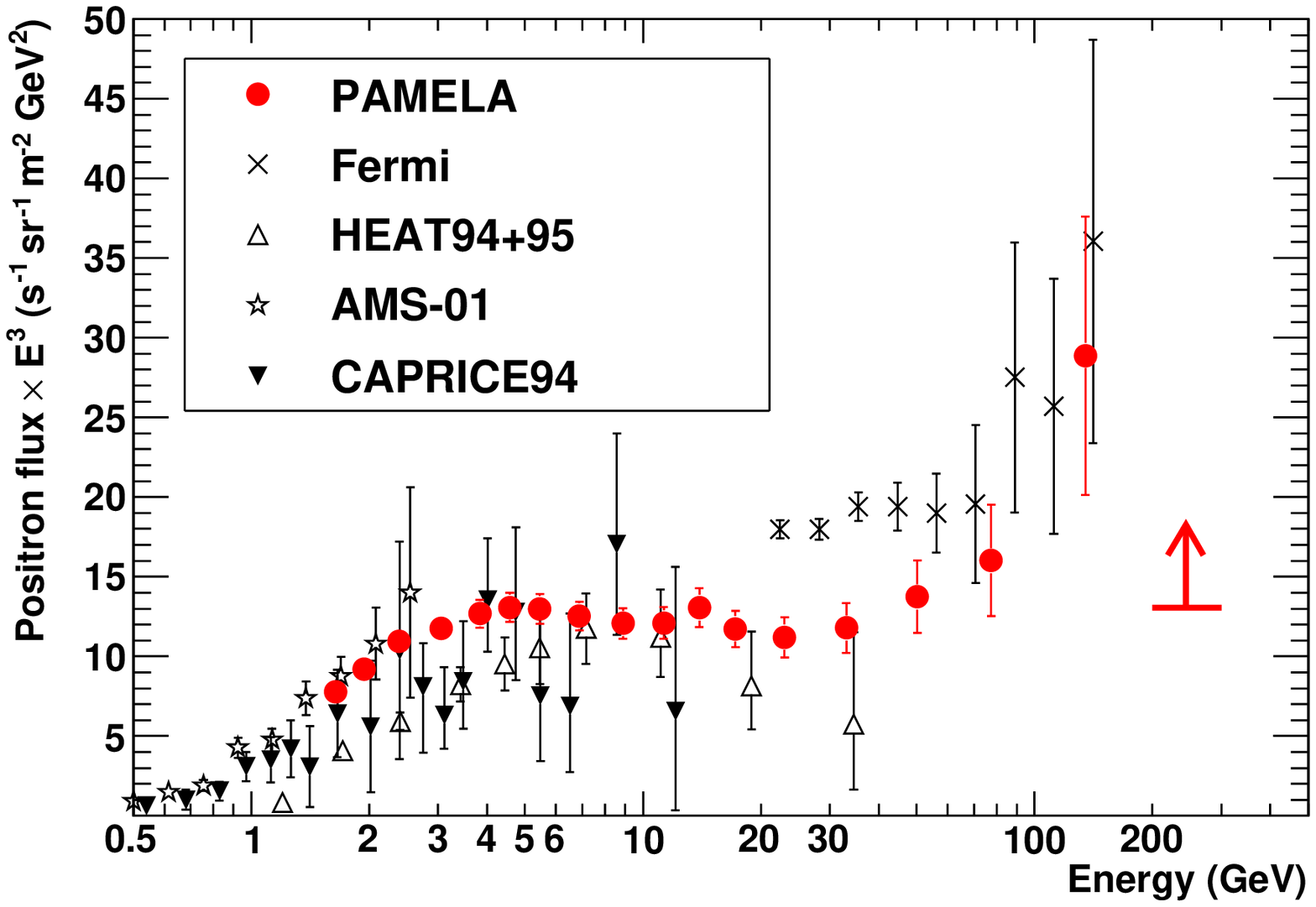} 
\end{center}
\caption{PAMELA and other recent
  measurements
of the 
positron energy spectrum:
CAPRICE94~\protect\cite{boe00}, HEAT94+95~\protect\cite{duv01},
AMS-01~\protect\cite{alc00b}, Fermi~\protect\cite{ack12a}. 
The PAMELA, Fermi and AMS-01 results are from 
space-borne experiments. The PAMELA lower limit is that for a 90\%
confidence level. PAMELA data points include both statistical and
systematic errors.
\label{fig:posflux1}}   
\end{figure}

\begin{figure}[ht]
\begin{center}
\includegraphics[width=\columnwidth]{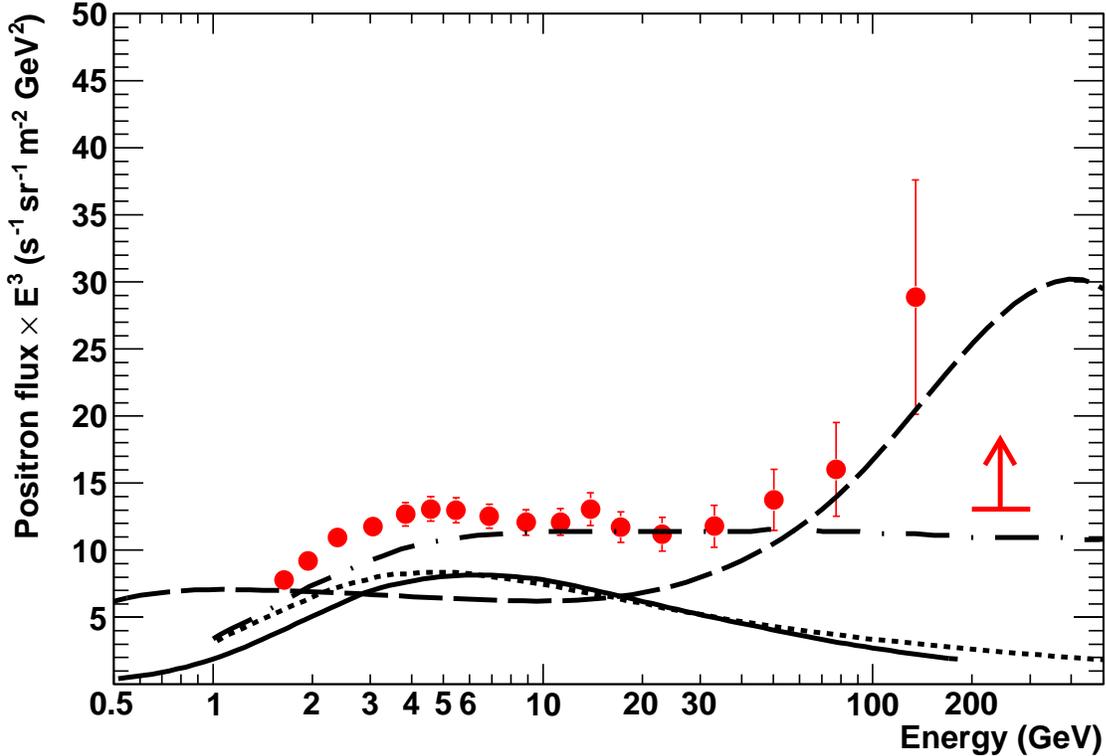}
\end{center}
\caption{The PAMELA positron energy spectrum data 
along with  theoretical calculations. 
The PAMELA lower limit is at the 90\% confidence level. PAMELA data
points include both statistical and 
systematic errors. 
The solid line shows the original GALPROP calculation 
\cite{str98} 
for pure secondary production  
of positrons during the propagation of cosmic rays in the Galaxy. 
The
dotted and dash-dotted lines show  
a recent calculation \cite{del10} for secondary and secondary plus
primary (from astrophysical sources)
positrons, 
respectively. The dashed line shows a calculation \cite{fin11} for
secondary plus primary positrons from dark matter annihilation.  
\label{fig:posflux2}}   
\end{figure}

\begin{figure}[ht]
\begin{center}
\includegraphics[width=\columnwidth]{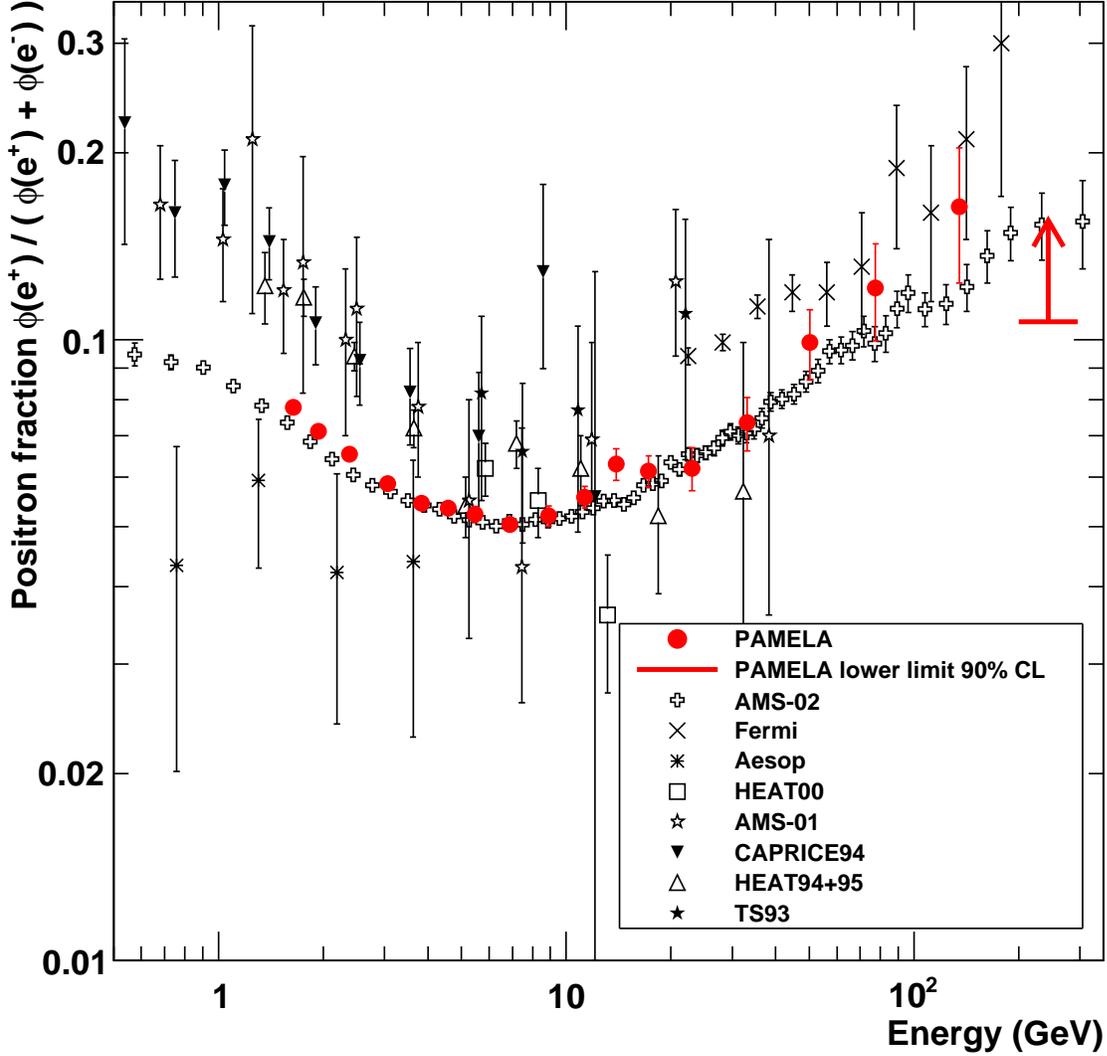}
\end{center}
\caption{PAMELA and other recent
  measurements of the
  positron fraction: TS93~\protect\cite{gol96},
  HEAT94+95~\protect\cite{bar97}, CAPRICE94~\protect\cite{boe00},
  AMS-01~\protect\cite{alc00b,agu07}, 
HEAT00~\protect\cite{bea04}, Aesop~\protect\cite{cle07},
  Fermi~\protect\cite{ack12a}, AMS-02~\protect\cite{agu13}.
The PAMELA, Fermi, AMS-01 and AMS-02 results are from 
space-borne experiments. 
\label{fig:posfrac1}}   
\end{figure}

The energy-binned positron data 
are given in Table~\ref{t:posi} and  
\begin{table}
\caption{Summary of positron 
results. 
The lower limit is that for a 90\% confidence level. 
For the flux the first and
second errors represent the statistical (68\% confidence level) and 
systematic uncertainties, respectively. \label{t:posi}}
\begin{ruledtabular}
\begin{tabular}{ccccc}
Rigidity     & Mean Kinetic & Observed                            &
Rescaled Flux           & \( \frac {\mbox{e$^+$}}{\mbox{(e$^+$ +
    e$^-$)}} \)  \\   
at the       & Energy at    & number of                           & at top of                                          & at top of \\  
spectrometer & top of       & events e$^+$                        & payload                                            & payload \\
GV/c         & payload GeV  &                                     & (GeV$^{-1}$s$^{-1}$sr$^{-1}$m$^{-2}$)$\times 10^{-3}$ &         \\ 
\hline
 1.5 - 1.8 & 1.64 & 4644 & $ 1762 \pm 24 \pm 111 $ & $ 0.0777 \pm 0.0011 $ \\
 1.8 - 2.1 & 1.94 & 3356 & $ 1262 \pm 21 \pm 80 $ & $ 0.0711 \pm 0.0012 $ \\
 2.1 - 2.7 & 2.38 & 2809 & $ 808 \pm 11 \pm 51 $ & $ 0.0653 \pm 0.0009 $ \\
 2.7 - 3.5 & 3.06 & 3755 & $ 411 \pm 6 \pm 26 $ & $ 0.0586 \pm 0.0010 $ \\
 3.5 - 4.2 & 3.83 & 3951 & $ 226 \pm 5 \pm 15 $ & $ 0.0545 \pm 0.0013 $ \\
 4.2 - 5 & 4.57 & 1520 & $ 137 \pm 3 \pm 9 $ & $ 0.0535 \pm 0.0014 $ \\
 5 - 6 & 5.46 & 1124 & $ 79.9 \pm 2.2 \pm 5.0 $ & $ 0.0523 \pm 0.0015 $ \\
 6 - 8 & 6.88 & 712 & $ 38.4 \pm 1.0 \pm 2.6 $ & $ 0.0504 \pm 0.0014 $ \\
 8 - 10 & 8.9 & 920 & $ 17.1 \pm 0.6 \pm 1.2 $ & $ 0.0520 \pm 0.0019 $ \\
 10 - 13 & 11.3 & 491 & $ 8.4 \pm 0.3 \pm 0.6 $ & $ 0.0557 \pm 0.0023 $ \\
 13 - 15 & 13.9 & 448 & $ 4.82 \pm 0.27 \pm 0.40 $ & $ 0.063 \pm 0.004 $ \\
 15 - 20 & 17.2 & 307 & $ 2.30 \pm 0.13 \pm 0.18 $ & $ 0.061 \pm 0.004 $ \\
 20 - 28 & 23 & 195 & $ 0.92 \pm 0.07 \pm 0.08 $ & $ 0.062 \pm 0.005 $ \\
 28 - 42 & 33.1 & 114 & $ 0.32 \pm 0.03 \pm 0.03 $ & $ 0.073 \pm 0.007 $ \\
 42 - 65 & 50.2 & 68 & $ 0.109 \pm 0.013 \pm 0.012 $ & $ 0.099 \pm 0.013 $ \\
 65 - 100 & 77.5 & 33 & $ 0.034 \pm 0.006 \pm 0.005 $ & $ 0.121 \pm 0.022 $ \\
 100 - 200 & 135 & 25 & $ 0.0118 \pm 0.0026 \pm 0.0024 $ & $ 0.163 \pm 0.040 $ \\
 200 - 300 &  &  & $ > 0.00091  $ & $ > 0.107   $ \\
\end{tabular}
\end{ruledtabular}
\end{table}
in Figures~\ref{fig:posflux1}, \ref{fig:posflux2} and \ref{fig:posfrac1} 
that show the
resulting positron energy spectrum~\footnote{The 
  fluxes are multiplied by E$^{3}$, where E is the energy in
  GeV. Reducing 
  the decades of variation of the flux, this
  allows for a clearer picture of the spectral shapes. However,
  this implies that the absolute energy uncertainties are convolved
  with the 
  flux uncertainties.} 
and positron fraction measured 
by PAMELA along with results from other recent 
cosmic-ray space- \cite{alc00b,agu07,ack12a,agu13} and 
balloon-borne \cite{boe00,duv01,gol96,bar97,cle07} experiments.
Between 200 GeV and 300 GeV only a lower limit at the 90\% confidence
level is  
presented because of a possible overestimation of the proton contamination
in the positron sample.  
The positron spectrum is significantly affected by solar modulation
below $\sim 10$~GeV. PAMELA data were acquired during a period of
slowly varying 
solar activity, close to a solar minimum, and the fluxes are averaged
over a three and a half year time period. The time dependence of the
low energy 
positron spectrum will be the topic of a future publication.  
In this 
paper only positron data above 1.5~GeV are presented. 
Taking into account the experimental
uncertainties and solar modulation effects,
the positron fraction presented here is in agreement  with the
previously published PAMELA results~\cite{adr09b,adr10a}. 

Figure
\ref{fig:posflux2} 
shows PAMELA data
along with theoretical 
predictions. 
The solid line shows the original GALPROP calculation 
\cite{str98} (calculated using the force field approximation
\cite{gle68} with solar modulation parameter $\Phi = 600$~MV)
assuming a pure 
secondary production of positrons during the propagation of
cosmic rays in the Galaxy. 
The dotted line shows a recent calculation of secondary positrons
\cite{del10}  
where it is argued that 
the progenitor proton flux is not
expected 
to vary significantly within a few kpc from Earth and so  
the flux of secondary positrons can be estimated with a few tens of
percent uncertainty, at most.
The high energy deviation of the
experimental data with respect to theoretical calculations has led to
many 
speculations about a primary origin of positrons. 
For example, the dashed line shows 
a calculation for secondary plus a primary contribution to the
positron flux 
resulting from annihilating dark matter particles of mass 1.2 TeV via
a dark 
gauge boson of mass 580 MeV to charged lepton pairs~\cite{fin11}.

A variety of astrophysical models have also been put 
forward to explain the positron excess.
Pulsars are well known particle accelerators.
Primary electrons are accelerated in the magnetosphere
of pulsars resulting in the emission of synchrotron gamma rays.
In the presence of the pulsar 
magnetic field, these gamma rays can produce positron and electron
pairs which escape into the interstellar medium after
$\sim 10^{5}$ years contributing to
the high-energy electron and positron cosmic-ray components
(e.g.,~\cite{ato95}). 
As an example, 
in Figure~\ref{fig:posflux2} the 
dash-dotted line indicates a contribution to the secondary component
from  
astrophysical sources such as pulsars~\cite{del10}. 
According to 
\cite{del10}, 
beyond 5-10 GeV 
there are poor constraints on the positron flux, e.g., from
radio observations. It should be noted that  
the contribution of primaries could take any shape
and that the dash-dotted line is just one possibility. Therefore, it has
been 
concluded \cite{del10} that the 
positron anomaly can be explained by a few prominent astrophysical
sources.
Furthermore,  the positron excess could also be explained
\cite{bla09b,fuj09,ahl09} by secondary production taking place in the
acceleration region of supernova
remnants.

Besides the positron excess at high energies, another feature is
clearly visible in the positron fraction  
data (Figure~\ref{fig:posfrac1}). At energies below
5~GeV, PAMELA results 
are systematically  
lower than other data
(except AMS-02 \cite{agu13} and Aesop data \cite{cle07}).
This low energy discrepancy with data  
collected during the 1990s, i.e., from the previous solar cycle
that favored positively-charged particles, is interpreted as a
consequence of charge-sign solar modulation effects \cite{pot01}.
The AMS-02 positron
fraction at low energies is, as expected, lower 
due to the increase in 
solar activity (e.g., see~\cite{mac13}) 
and shows the same high energy rise.
This agreement gives good confidence that
the increase of the positron flux can be ascribed to a physical
effect and not to systematics affecting the measurements.

\begin{acknowledgments}
We acknowledge support from The Italian Space Agency 
(ASI), Deutsches Zentrum f\"{u}r Luft- und Raumfahrt (DLR), The
Swedish National Space  
Board, The Swedish Research Council, The Russian Space Agency
(Roscosmos) and The
Russian Foundation for Basic Research.  
\end{acknowledgments}

\bibliography{pamela_psflux}

\end{document}